\documentclass[aps,prl,twocolumn,superscriptaddress,showpacs,preprintnumbers,amsmath,amssymb]{revtex4}


\usepackage{dcolumn}

\input epsf
\usepackage{tabularx}
\usepackage{ulem}
\usepackage{epsfig}
\usepackage{verbatim}
\usepackage{subfigure}
\usepackage{graphicx}
\usepackage{amssymb}
\usepackage{epstopdf}
\usepackage{color}
\usepackage{xspace}
\usepackage[amssymb,pstricks,Gray,squaren]{SIunits}
\usepackage{header}
\usepackage{graphics}
\usepackage{color}

\RequirePackage{relsize} 
\begin{document}

\title{{\boldmath 
\begin{flushright}
{\small UCHEP-08-04 }\\
\end{flushright}
\vskip0.10in
Evidence for Neutral \B Meson Decays to \OmeKstz }}

\affiliation{Budker Institute of Nuclear Physics, Novosibirsk}
\affiliation{University of Cincinnati, Cincinnati, Ohio 45221}
\affiliation{T. Ko\'{s}ciuszko Cracow University of Technology, Krakow}
\affiliation{Justus-Liebig-Universit\"at Gie\ss{}en, Gie\ss{}en}
\affiliation{The Graduate University for Advanced Studies, Hayama}
\affiliation{Hanyang University, Seoul}
\affiliation{University of Hawaii, Honolulu, Hawaii 96822}
\affiliation{High Energy Accelerator Research Organization (KEK), Tsukuba}
\affiliation{University of Illinois at Urbana-Champaign, Urbana, Illinois 61801}
\affiliation{Institute of High Energy Physics, Chinese Academy of Sciences, Beijing}
\affiliation{Institute of High Energy Physics, Vienna}
\affiliation{Institute of High Energy Physics, Protvino}
\affiliation{Institute for Theoretical and Experimental Physics, Moscow}
\affiliation{J. Stefan Institute, Ljubljana}
\affiliation{Kanagawa University, Yokohama}
\affiliation{Korea University, Seoul}
\affiliation{Kyungpook National University, Taegu}
\affiliation{\'Ecole Polytechnique F\'ed\'erale de Lausanne (EPFL), Lausanne}
\affiliation{Faculty of Mathematics and Physics, University of Ljubljana, Ljubljana}
\affiliation{University of Maribor, Maribor}
\affiliation{University of Melbourne, School of Physics, Victoria 3010}
\affiliation{Nagoya University, Nagoya}
\affiliation{Nara Women's University, Nara}
\affiliation{National Central University, Chung-li}
\affiliation{National United University, Miao Li}
\affiliation{Department of Physics, National Taiwan University, Taipei}
\affiliation{H. Niewodniczanski Institute of Nuclear Physics, Krakow}
\affiliation{Nippon Dental University, Niigata}
\affiliation{Niigata University, Niigata}
\affiliation{University of Nova Gorica, Nova Gorica}
\affiliation{Osaka City University, Osaka}
\affiliation{Osaka University, Osaka}
\affiliation{Panjab University, Chandigarh}
\affiliation{RIKEN BNL Research Center, Upton, New York 11973}
\affiliation{Saga University, Saga}
\affiliation{University of Science and Technology of China, Hefei}
\affiliation{Seoul National University, Seoul}
\affiliation{Sungkyunkwan University, Suwon}
\affiliation{University of Sydney, Sydney, New South Wales}
\affiliation{Toho University, Funabashi}
\affiliation{Tohoku Gakuin University, Tagajo}
\affiliation{Tohoku University, Sendai}
\affiliation{Department of Physics, University of Tokyo, Tokyo}
\affiliation{Tokyo Institute of Technology, Tokyo}
\affiliation{Tokyo Metropolitan University, Tokyo}
\affiliation{Tokyo University of Agriculture and Technology, Tokyo}
\affiliation{Virginia Polytechnic Institute and State University, Blacksburg, Virginia 24061}
\affiliation{Yonsei University, Seoul}
\author{P.~Goldenzweig}\affiliation{University of Cincinnati, Cincinnati, Ohio 45221} 
\author{A.~J.~Schwartz}\affiliation{University of Cincinnati, Cincinnati, Ohio 45221} 
\author{I.~Adachi}\affiliation{High Energy Accelerator Research Organization (KEK), Tsukuba} 
\author{H.~Aihara}\affiliation{Department of Physics, University of Tokyo, Tokyo} 
\author{K.~Arinstein}\affiliation{Budker Institute of Nuclear Physics, Novosibirsk} 
\author{V.~Aulchenko}\affiliation{Budker Institute of Nuclear Physics, Novosibirsk} 
\author{T.~Aushev}\affiliation{\'Ecole Polytechnique F\'ed\'erale de Lausanne (EPFL), Lausanne}\affiliation{Institute for Theoretical and Experimental Physics, Moscow} 
\author{S.~Bahinipati}\affiliation{University of Cincinnati, Cincinnati, Ohio 45221} 
\author{A.~M.~Bakich}\affiliation{University of Sydney, Sydney, New South Wales} 
  \author{A.~Bay}\affiliation{\'Ecole Polytechnique F\'ed\'erale de Lausanne (EPFL), Lausanne} 
% ...
\collaboration{The Belle Collaboration}


\begin{abstract}
We present the results of a study of the charmless vector-vector decay \BzToOmeKstz with \NBBbarVal \BBbar pairs collected with the Belle detector at the KEKB \ee collider. We measure the branching fraction to be \BfBzToOmeKstzValErrTxt with \OmeKstzSigSystCorr significance. We also perform a helicity analysis of the \Ome and \Kstz vector mesons, and obtain the longitudinal polarization fraction \fLResult. Finally, we measure a large non-resonant branching fraction \BfBzToOmeKpPimValErrNoTextRed with a significance of \OmeKpPimSigSystCorr.
\end{abstract}

\maketitle

The study of branching fractions and angular distributions of \B meson decays to hadronic final states tests our understanding of both weak and strong interactions. Recently, \B decays mediated by \bsqq penguin amplitudes have received much attention in the literature. Unlike \bc spectator amplitudes (which are much better measured), penguin amplitudes contain an internal loop and thus are potentially sensitive to new propagators and couplings. Such mediating particles may have an energy scale too high to access directly. Several measured \bsqq decays have yielded unexpected results; e.g., the decays \BToPhiKst and \BToRhoKstz are found to have large transverse polarization~\cite{phiKstz_rhoKstz}, and $B$ decays to the closely related final states \KpPim and \KpPiz exhibit different patterns of direct $CP$ violation~\cite{kpiDCPV}. These results are difficult to accommodate within the Standard Model and may indicate the presence of new physics~\cite{low_fL_theory}. Furthermore, \bsqq decays are useful for determining the angles ${\phi_{2}}$ and ${\phi_{3}}$ of the unitarity triangle~\cite{phi3}.

In this Letter we present a study of the \bsdd decay \BzToOmeKstz. Theoretical calculations for the branching fraction cover the range \TheoryBrRange~\cite{br_theory}. Previously, this mode has been searched for by CLEO~\cite{cleo_wKstz} and BaBar~\cite{babar_wKstz}; the latter group observed an excess of events with a significance of 2.4\sig. Our analysis uses \dataset of data containing \NBBbarVal \BBbar pairs; this sample is almost three times larger than that used in Ref.~\cite{babar_wKstz}. With this large data set we are able to measure both the branching fraction and longitudinal polarization fraction for \BzToOmeKstz, and the branching fraction for non-resonant \BzToOmeKpPim. The data were collected with the Belle detector~\cite{belle_detector} at the KEKB~\cite{kekb} \ee asymmetric-energy (3.5 GeV on 8.0 GeV) collider with a center-of-mass (CM) energy at the \ups resonance. The production rates of \BzBzbar and \BpBm pairs are assumed to be equal. 

The Belle detector is a large-solid-angle spectrometer. It includes a silicon vertex detector (SVD), a 50-layer central drift chamber (CDC), an array of aerogel threshold  Cherenkov counters (ACC), time-of-flight scintillation counters (TOF), and an electromagnetic calorimeter (ECL) comprised of CsI(Tl) crystals located inside a superconducting solenoid coil that provides a 1.5 T magnetic field.

The \B-daughter candidates are reconstructed through the decays \decayOme, \decayKstz and \decaypiz~\cite{charge_conjugate_text}. A charged track is identified as a pion or kaon by using particle identification (PID) information from the CDC, ACC and TOF systems. We reduce the number of poor quality tracks by requiring that ${|dz| < 4.0~\cm}$ and ${dr < 0.2~\cm}$, where ${|dz|}$ and ${dr}$ are the distances of closest approach of a track to the interaction point along the $z$-axis (opposite the direction of the positron beam) and in the transverse plane, respectively. In addition, we require that each charged track have a transverse momentum ${p_{T} > 0.1~\gevc}$ and a minimum number of SVD hits. Tracks matched with clusters in the ECL that are consistent with an electron hypothesis are rejected.

Photons used for \piz reconstruction are required to have energies in the laboratory frame greater than 50 (100) MeV for the ECL barrel (endcap), which subtends \ECLbarrelPRL (\ECLendcapLowPRL and \ECLendcapHighPRL) with respect to the beam axis. We require \piz candidates to have an invariant mass in the range \pizcutPRL ($\pm3\sig$ in \mpizPDG resolution) and a momentum in the laboratory frame \pizmomcut.  

We select \Ome mesons with an invariant mass in the range \OmefitPRL, and \Kstz mesons with \KstzfitPRL. These windows include sideband regions to provide discrimination between signal and background components in the maximum-likelihood (ML) fit described below. To reduce combinatorial background arising from low-momentum kaons and pions, we require that \CosKstzCut, where \ThetaKstz is the \Kstz helicity angle defined as the angle between the direction of the \kp and the direction opposite to the \Bz momentum in the \Kstz rest frame. 

Signal decays are identified using the energy difference (\de) and the beam-energy-constrained mass (\mbc). These are defined as \defde and \defmbc, where \ebeam denotes the beam energy and \eb and \pb denote the energy and momentum, respectively, of the candidate \B-meson,  all evaluated in the \ee CM frame. We retain events satisfying \defitPRL and \mbcfitPRL, and define a signal region \desigPRL, \mbcsigPRL.

The dominant source of background is continuum \eeqq (${q = u,d,s,c}$) production. To discriminate relatively spherical \BBbar events from jet-like \qqbar events, we use 16 modified Fox-Wolfram moments (combined into a Fisher discriminant \F~\cite{fisher}), the CM polar angle between the \B direction and the $z$-axis (\thetab), and the displacement along the $z$-axis between the signal \B vertex and that of the other \B in the event (\deltaz). Further discrimination is provided by a ${b}$-flavor tagging algorithm~\cite{tag_algorithm}, which identifies the flavor of the \B meson accompanying the signal candidate via its decay products: charged leptons, kaons, and $\Lambda$'s. This algorithm yields a quality factor \tagr, which ranges from zero for no flavor discrimination to one for unambiguous flavor assignment. 

We use Monte Carlo (MC) simulated signal~\cite{GEANT} and data sideband events [defined as \mbcsbdPRL, \defitPRL] to obtain probability density functions (PDFs) for \F, \costhetab and \deltaz. These are multiplied together to form signal (\Lsig) and \qqbar background (\Lqq) likelihood functions, and we require that \Lratiodef be above a threshold.  We divide the events into six bins of \tagr and determine the optimum \Lratio threshold for each bin by maximizing a figure-of-merit \fomdefPRL, where \FOMs (\FOMb) is the number of signal (background) events in the signal region. This optimization rejects 99\% of the \qqbar background while preserving 50\% of the signal.

The fraction of events having multiple candidates is 12\%. We choose the candidate in an event to be the one that minimizes the quantity \bcs. From MC studies we find that this choice selects the correct candidate 90\% of the time. We also find that \fSCFOmeKstzPaperVal of signal decays have at least one particle incorrectly identified but pass all selection criteria; these are referred to as ``self-cross-feed" (SCF) events.


We obtain the yields using a four-dimensional (4D) extended unbinned ML fit to \de, \mbc, \mOme and \mKstz. The likelihood function is given by
\begin{equation} \label{ll def}
\LFdefinition,
\end{equation}
where ${Y_{j}}$ is the yield of component ${j}$, ${{\cal{P}}^{i}_{j}}$ is the PDF for component ${j}$, and $i$ runs over all events in the sample. We include PDFs for the signal, \qqbar background (\qqbar), charm \B-decay background (\bc), charmless \B-decay background (\bsud), and non-resonant \BzToOmeKpPim decays. The MC acceptances for non-resonant \BzToKstzPpPimPiz and \BzToKpPimPpPimPiz are negligibly small and thus we do not consider these channels.

The PDF for each component is defined as \PDFdefinition. For the signal and \OmeKpPim components, we split the PDFs into two parts: \PDFdefinitionSCF, where \fSCF is the SCF fraction (\fSCFOmeKpPimPaperVal for \OmeKpPim), and ``true" represents the correctly reconstructed decays. For the \qqbar, \bc and \bsud backgrounds, no sizable correlations are found among the fitted variables. For the signal and \OmeKpPim components, there are small correlations that are accounted for as described below.

The \Kstz and \Ome resonances are modeled with Breit-Wigner functions whose widths are fixed to their PDG~\cite{PDG2006} values. The Breit-Wigner function used to describe the \Ome resonance is convolved with a Gaussian of ${\sigma = 5.7~\mathrm{MeV}}$ to take into account the detector resolution. This value, along with the means for both resonances and the fraction of \qqbar background events containing \Ome's and \Kstz's, are obtained from fitting the \mKstz and \mOme spectra of events in the data sideband.

All other PDF shapes are obtained from MC simulation. For the signal and \OmeKpPim PDFs, the sum of a Crystal Ball line shape~\cite{cbls} and Gaussian is used to describe \de, and the sum of two Gaussians is used to describe \mbc. To take into account small differences between the MC simulations and data, the \mbc and \de shapes for the signal and \OmeKpPim PDFs are corrected according to calibration factors determined from a large \bztodmrhop, \dmtokpp control sample. The \mKstz PDF for \OmeKpPim decays is represented by a threshold function with parameters determined from MC events where the \Kpi final state is distributed uniformly over phase space. 

For the \qqbar background, we use a threshold ARGUS~\cite{argus} function to describe \mbc, and linear functions to describe \de and the combinatorial shapes of \mOme and \mKstz. The \mbc and \de shapes of the \bc background are described by an ARGUS function and a second-order Chebyshev polynomial, respectively. The remaining PDF shapes are modeled with non-parametric PDFs using Kernel Estimation~\cite{keys}.

The following parameters vary in our final fit to the data: the signal, \OmeKpPim, \bc and \qqbar yields, and the \qqbar PDF parameters describing the \de, \mbc and combinatorial shapes of \mOme and \mKstz. The fraction of \bsud events (\fbsud) is small (\fbsudVal) and fixed to the MC value. The \fSCF for signal and \OmeKpPim decays are also fixed to their MC values.

\begin{figure}[t]
\mbox{\epsfxsize=1.68in \epsfbox{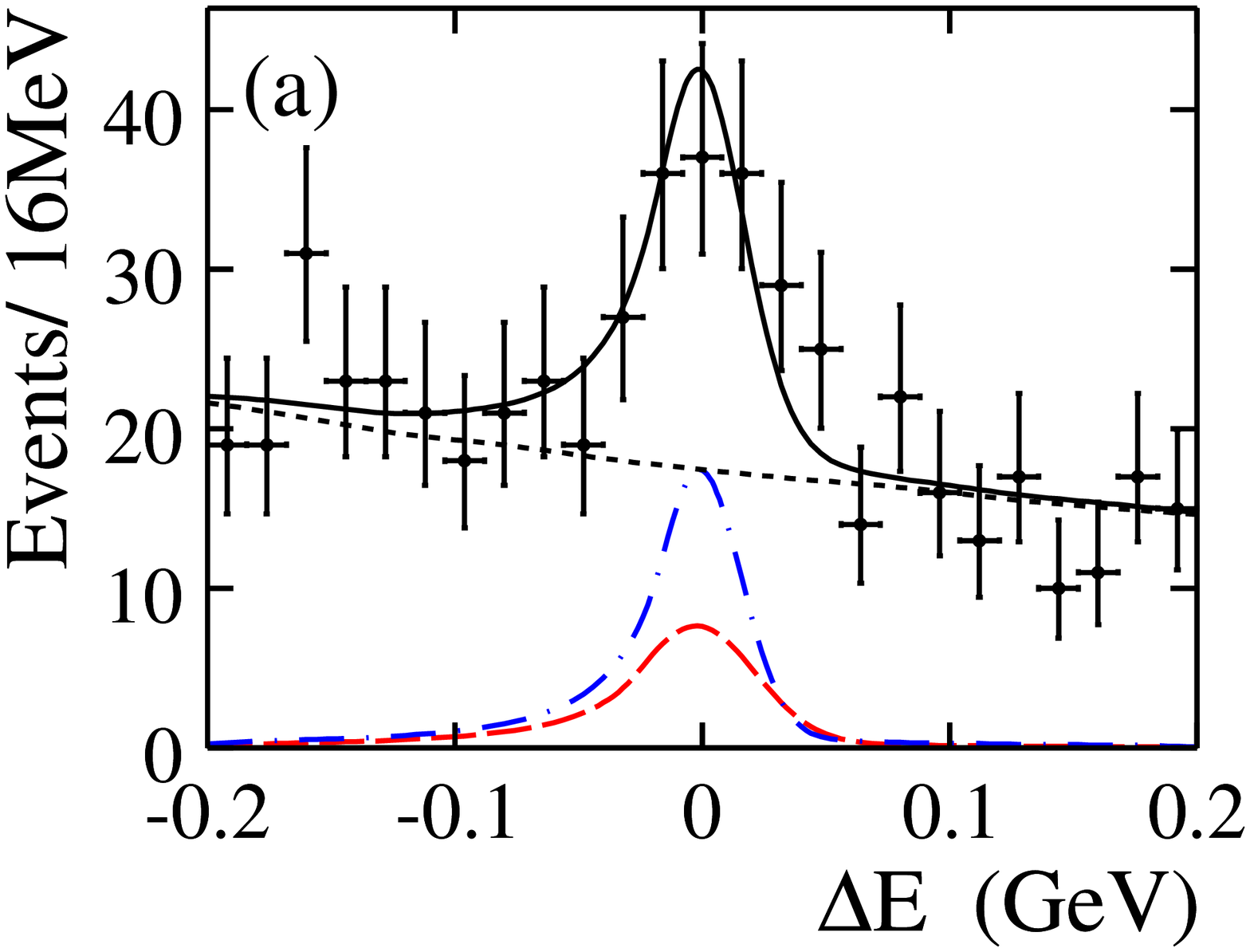}}
\mbox{\epsfxsize=1.68in \epsfbox{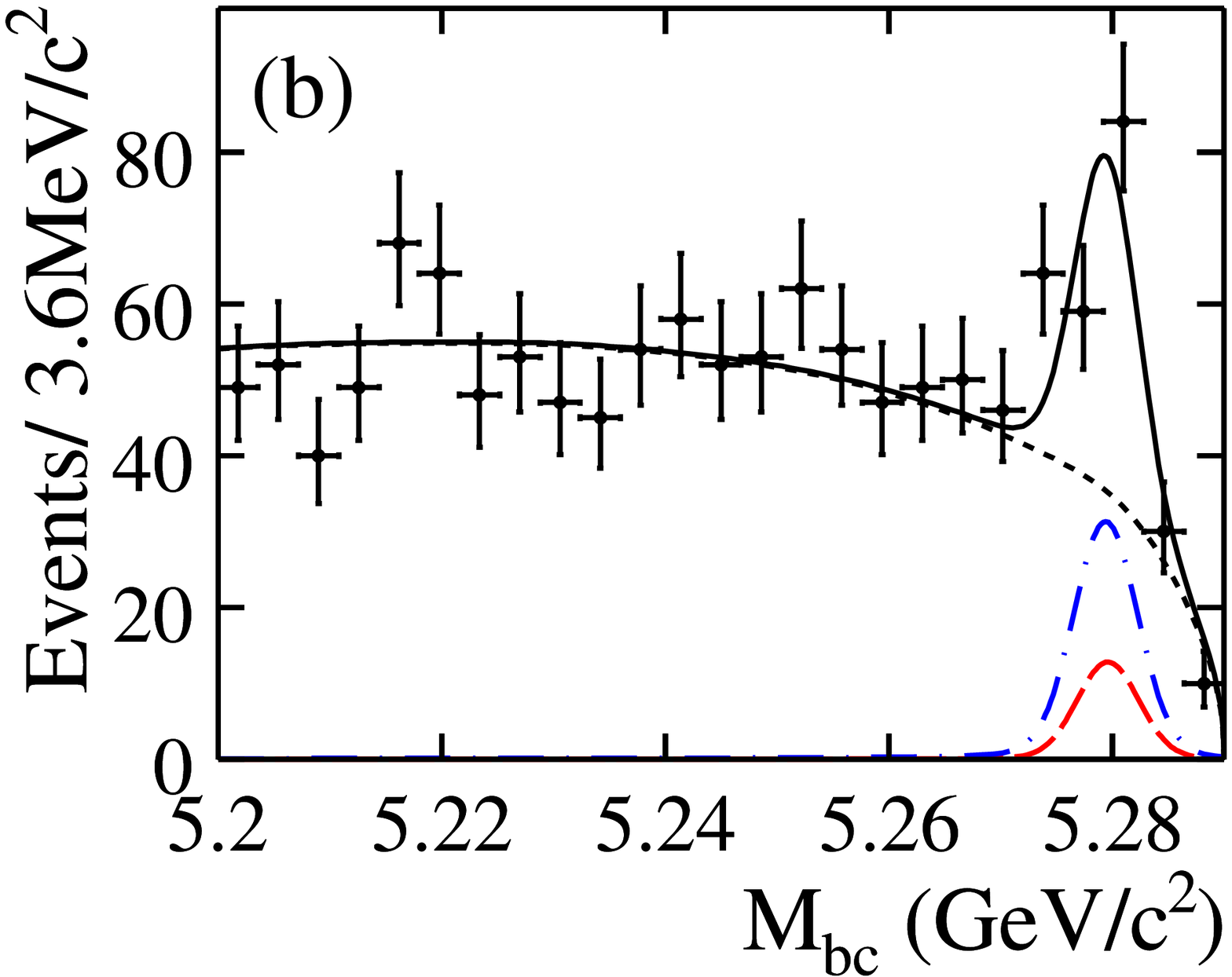}}\\
\mbox{\epsfxsize=1.68in \epsfbox{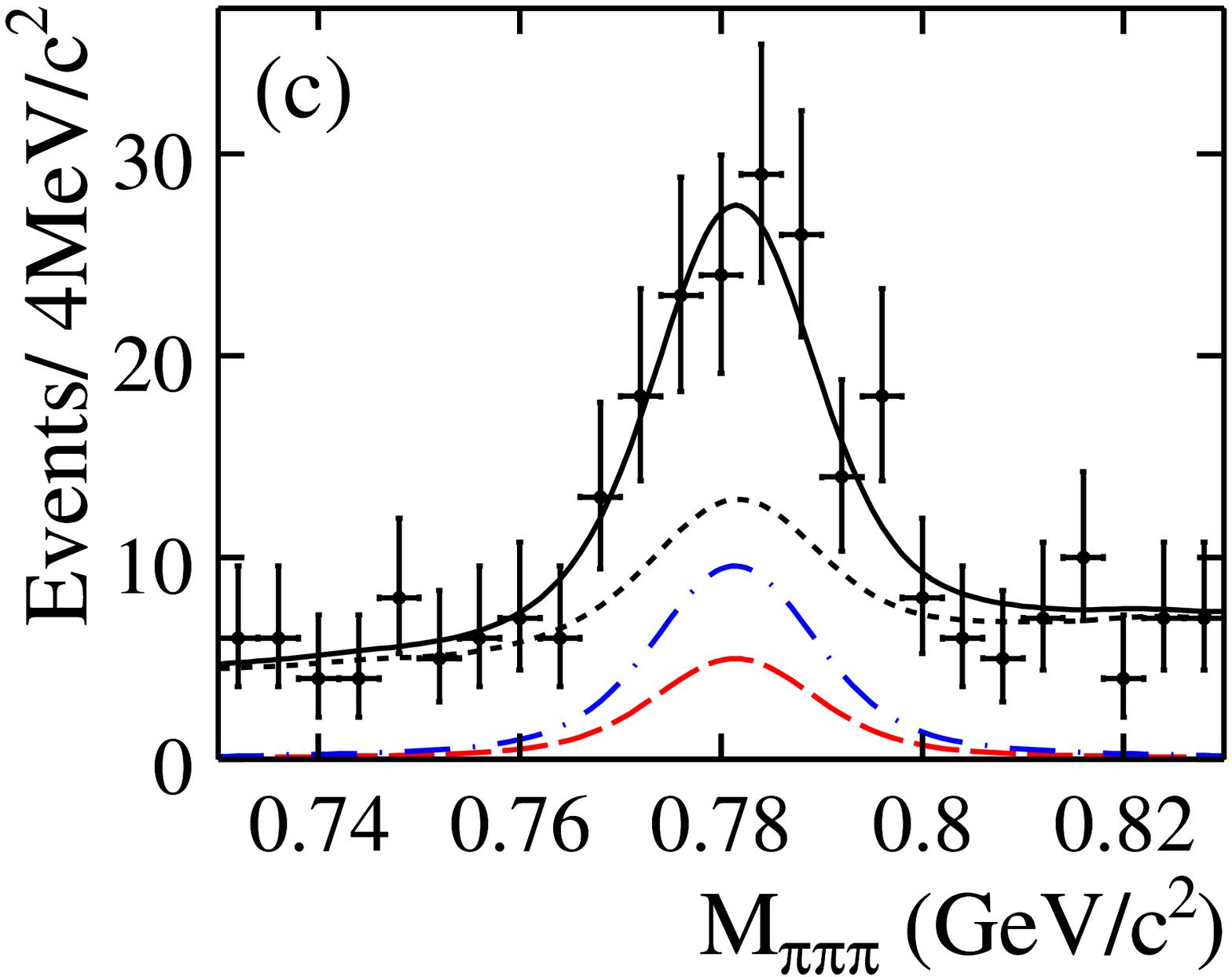}}
\mbox{\epsfxsize=1.68in \epsfbox{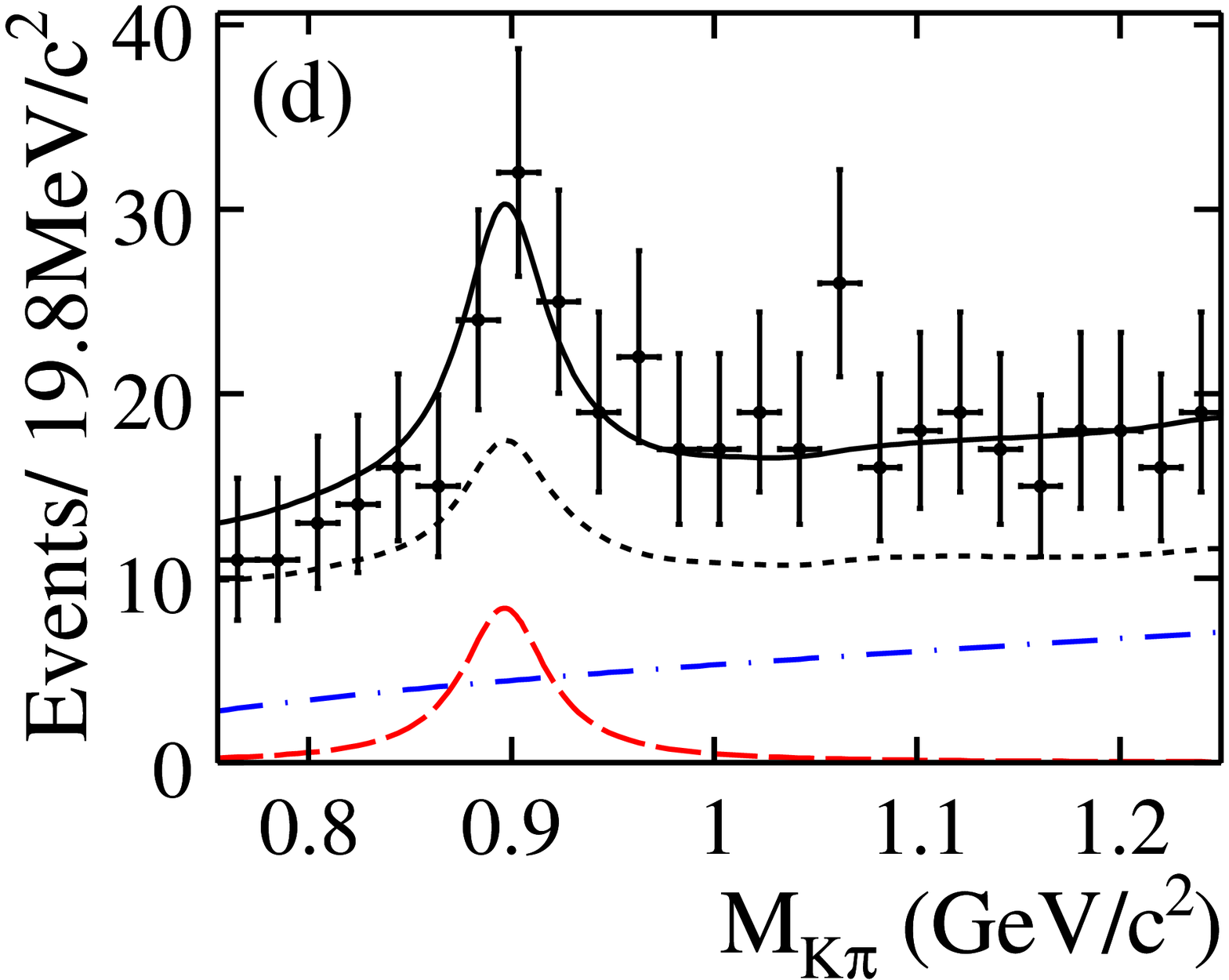}}
\caption{\label{4d fit} Projections of the fit results onto (a) \de, (b) \mbc, (c) \mOme and (d) \mKstz for candidates satisfying (except for the variable plotted) the criteria \desigPRL, \mbcsigPRL and \KstzsigPRL. The curves are for \OmeKstz (dashed), \OmeKpPim (dot-dashed), the sum of the backgrounds (dotted), and the total (solid).}
\end{figure}

The fit results are listed in Table \ref{br summary table} and the projections are shown in Fig. \ref{4d fit}. With the fitted yields $Y$, we calculate the branching fraction \cb as \BfDefPaper, where \epsMC is the event selection efficiency including daughter branching fractions obtained from MC simulation, \NBBbar is the number of \BBbar pairs produced, and \PIDeffDef is an efficiency correction for the charged track selection that takes into account small differences between MC values and data. Our signal MC simulation is generated with \fL = 0.5; the change in acceptance for other values of \fL is taken as a systematic error. For \OmeKpPim, \epsMC and \calB are for \KstzfitPRL. The significance is defined as \SqrtNegTwoLnLzLm, where \Lmax (\Lzero) is the value of the likelihood function when the yield is allowed to vary (set to 0). The systematic uncertainty is included by convolving the likelihood function with a Gaussian whose width is equal to the systematic error. For signal and \OmeKpPim decays, we account for small correlations between the fitted variables by fitting ensembles of simulated experiments containing all signal and background components. The correlations give rise to biases of \OmeKstzBiasValPRL and \OmeKpPimBiasValPRL events for signal and \OmeKpPim, respectively. We correct the fitted yields for these biases.   
  
\begin{table}[t]
\small
\begin{center}
\caption{\label{br summary table} Signal yield $Y$ and its statistical uncertainty, MC efficiency \epsMC, PID efficiency \PIDeffDef, significance $S$ with systematic uncertainties included, and measured branching fraction ${\cal B}$. For \OmeKpPim, \epsMC and \calB are for \KstzfitPRL. For \calB, the first (second) error is statistical (systematic).}
\begin{tabular*}{0.48\textwidth}{@{\extracolsep{\fill}}lccccc}
\hline \hline
Mode & $Y$  & \epsMC(\%)  & \PIDeffDef & $S$ & ${\cal B}$  ($10^{-6}$)  \\ 
\hline
\OmeKstz      &  \OmeKstzFourDFitResErrCorrVal     &\OmeKstzEffXBrProdVal  & \PIDeffVal   & 
\OmeKstzSigSystCorrVal   &  \BfBzToOmeKstzValErr   \\ \hline  
\OmeKpPim  &  \OmeKpPimFourDFitResErrCorrVal & \OmeKpPimEffXBrProdValRed & \PIDeffVal & 
\OmeKpPimSigSystCorrVal  &  \BfBzToOmeKpPimValErrRed  \\ \hline  \hline
\end{tabular*}
\end{center}
\end{table}

\begin{figure}[b]
\mbox{\epsfxsize=3.4in \epsfbox{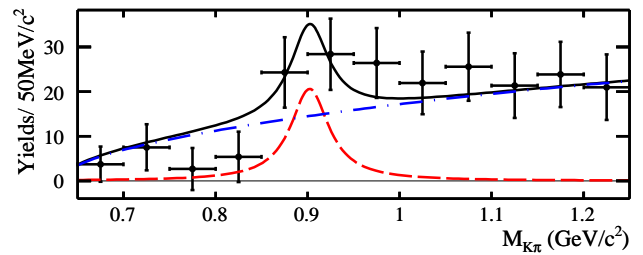}}
\caption{\label{bkgd sub fit}Signal + \OmeKpPim yields obtained from 2D fits to \de and \mbc in bins of \mKstz. The curves are for \OmeKstz (dashed), \OmeKpPim (dot-dashed), and the total (solid).}
\end{figure}

The main sources of systematic error are: track reconstruction efficiency (\ErrTrackPRL per track); \piz efficiency (\ErrPizPRL); PID (\ErrPIDPRL); \NBBbar (\ErrNBBbarPRL); MC statistics (\ErrMCOmeKstzPRL); PDF shapes (\ErrPDFOmeKstzPosNeg); \fbsud (\ErrfbsudOmeKstzPRL); \fSCF (\ErrfSCFOmeKstzPosNeg); the \de fit range (\ErrDeRangeOmeKstzPosNeg); fitting bias (\ErrfFitBiasOmeKstzPosNeg); the effect of higher \Kstz resonances (\ErrfHighKstzOmeKstzPosNeg); \fL (\ErrfLOmeKstzPosNeg); and \Lratio (\ErrLqqPRL). The errors on the PDF shapes are obtained by varying all fixed parameters by ${\pm1\sigma}$ and taking the fractional change in the yield as the systematic error. To obtain the error due to \fSCF and \fbsud, we vary these fractions by $\pm50\%$. The uncertainty in the yield bias correction is taken to be the sum in quadrature of the statistical uncertainty on the correction and half the correction value. We consider the effects of higher \Kstz resonances by including a PDF for \BzToOmeKstzHigh and repeating the 4D fit with the \OmeKstzhigh yield fixed to the value obtained by extrapolating from a higher \mKstz region. The error due to the uncertainty in \fL is obtained by varying \fL by its errors measured below. To obtain the uncertainty due to the \Lratio requirement, we vary the \Lratio thresholds, and we also calculate the data/MC efficiency ratio for the \bztodmrhop, \dmtokpp control sample. 

We study the effects of interference between \BzToOmeKstz and \BzToOmeKpPim decays as follows. We modify the Breit-Wigner PDF describing the \Kstz resonance of the signal to include an interfering amplitude and phase; for lack of more information, we take this amplitude to be constant in \mKstz. We uniformly vary the amplitude and phase from zero to a maximum and, for each case, generate and fit a large ensemble of toy MC experiments. The rms spread of deviations about the true value is taken as the systematic error (\ErrIntOmeKstzPosNeg for \OmeKstz). Combining all errors in quadrature gives a total systematic error of (\ErrSumOmeKstzPosNeg). The systematic errors considered for \OmeKpPim are similar; the total is (\ErrSumOmeKpPimPosNeg).

To verify the large \OmeKpPim contribution (see Table~\ref{br summary table}), we bin the data in \mKstz from \bkgdsubKstzfitPRL and, for each bin, perform a two-dimensional (2D) fit to \de and \mbc. The likelihood function consists of three components: signal + \OmeKpPim, \qqbar + \bc, and \bsud. We plot the resulting yields of signal + \OmeKpPim as a function of \mKstz (Fig. \ref{bkgd sub fit}) and fit this distribution to extract the signal and \OmeKpPim components. For \KstzfitPRL we obtain yields of \OmeKstzTwoDFitResErrVal and \OmeKpPimTwoDFitResErrVal for signal and \OmeKpPim, respectively; these values are in good agreement with the results of the 4D fit after accounting for the fit bias.

\begin{figure}[t]
\mbox{\epsfxsize=1.68in \epsfbox{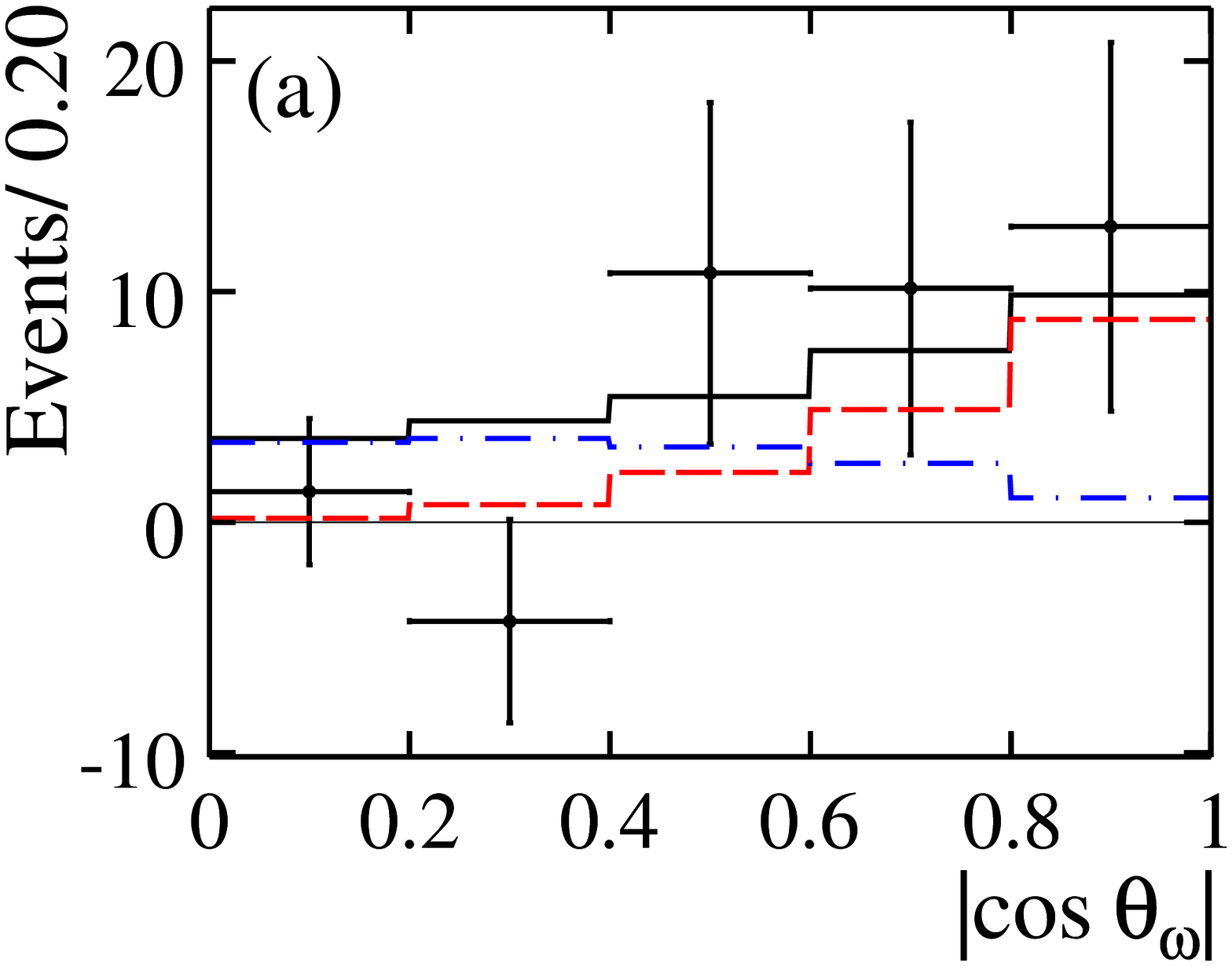}}
\mbox{\epsfxsize=1.68in \epsfbox{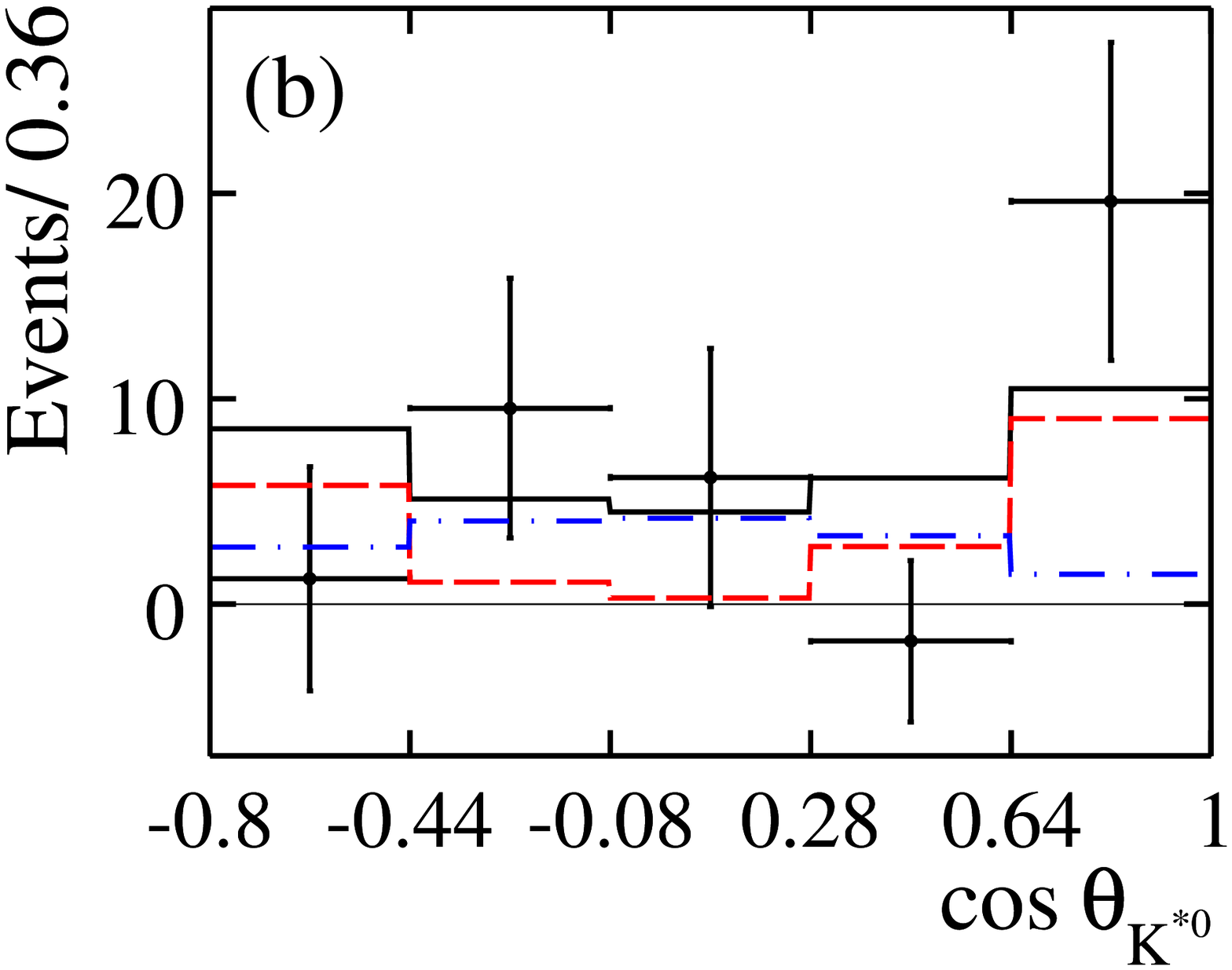}}
\caption[]{Signal yields obtained from 4D fits to \de, \mbc, \mKstz, and \mOme, in bins of (a) \AbsCosOme and (b) \CosKstz. The solid histograms show the results of the simultaneous \chisq fit. The dashed (dot-dashed) histograms are the \Az (\Apm) component.  \label{helicity cosines cv}}
\end{figure}

The differential decay width, after integrating over the angle between the decay planes of the \Ome and \Kstz mesons, is proportional to \HelDefFinalOmeKstzPropRHSNoFrac. Here, \ThetaOme is the \Ome helicity angle defined as the angle between the normal to the three-pion decay plane and the negative of the \Bz momentum in the \Ome rest frame. The fraction of longitudinal polarization \fLDefPRL, where \Alambda are the helicity amplitudes for the longitudinal (\LambdaZ) and transverse (\LambdaPM) states~\cite{hel_MC_dist}. To determine \fL, we bin the data in \AbsCosOme and \CosKstz and, for each bin, perform a 4D fit to \de, \mbc, \mKstz, and \mOme. The resulting signal yields as a function of the helicity cosines are shown in Fig. \ref{helicity cosines cv}. We perform a simultaneous \chisq fit to these distributions, where the only floating parameter is \fL. The PDFs for the \Az and \Apm helicity states are determined from MC simulation to take into account the detection efficiency. The statistical error is obtained from a toy MC study (the rms spread of the residuals from a large ensemble), since the errors in the distributions of Fig. \ref{helicity cosines cv} are correlated. Using a large toy MC sample we measure a 2\% bias in the fitting procedure, which we use to correct the central value. 

There are six main sources of systematic error in \fL: uncertainty in the PDF shapes \fLSystPDFShape; the fractions \fbsud \fLSystfbsud and \fSCF \fLSystfSCF; fitting bias \fLSystBias; interference \fLSystInterference; and the \Lratio requirement \fLSystRqq. Adding the various systematic contributions in quadrature, we obtain a longitudinal polarization fraction
\begin{equation}
\fLResult.
\end{equation}

In summary, using \NBBbarVal \BBbar pairs we have found evidence for the \BzToOmeKstz decay with a significance of \OmeKstzSigSystCorr. We measure the branching fraction to be \BfBzToOmeKstzValErrTxt. Our result is in agreement with theoretical estimates~\cite{br_theory}, and with the central value obtained by BaBar~\cite{babar_wKstz}. We also perform a helicity analysis of the \Ome and \Kstz vector mesons and measure a longitudinal polarization fraction \fLResult. This central value is lower than that predicted by most theoretical models but is similar to that measured for other \bsqq decays~\cite{phiKstz_rhoKstz}. In addition, we measure a large non-resonant branching fraction \BfBzToOmeKpPimValErrNoTextRed with a significance of \OmeKpPimSigSystCorr. Assuming a uniform phase space distribution, this implies a branching fraction of \BzToOmeKpPimValErrStatSyst over the whole region.

We thank the KEKB group for excellent operation of the
accelerator, the KEK cryogenics group for efficient solenoid
operations, and the KEK computer group and
the NII for valuable computing and SINET3 network
support.  We acknowledge support from MEXT and JSPS (Japan);
ARC and DEST (Australia); NSFC (China); 
DST (India); MOEHRD, KOSEF and KRF (Korea); 
KBN (Poland); MES and RFAAE (Russia); ARRS (Slovenia); SNSF (Switzerland); 
NSC and MOE (Taiwan); and DOE (USA).



\begin{thebibliography}{99}
\bibitem{charge_conjugate_text}
Charge-conjugate decays are included unless explicitly stated otherwise.
\end{thebibliography}
\end{document}